\documentclass[a4paper,11pt,dvipsnames]{article}
\usepackage{pos}

\usepackage{amsmath}
\usepackage{cancel}
\usepackage{epstopdf}
\usepackage{mathtools}
\usepackage{bbm}
\usepackage{faktor}
\usepackage{tensor}
\usepackage{bbm}
\usepackage{slashed}

\usepackage{graphicx}
\usepackage[export]{adjustbox}
\usepackage[]{xcolor}
\usepackage{tikz}
\usepackage[compat=1.1.0]{tikz-feynman}
\tikzfeynmanset{/tikzfeynman/warn luatex = false} 
\usetikzlibrary{calc}

\usepackage{subcaption} 

\usepackage{spinors}
\usepackage{auxiliary_definitions}

\def\TTr{\mathop{\rm ITr}\nolimits}
\def\modout#1#2{{#1}/{#2}}
\def\sufrac#1#2{{#1}/{#2}}

\def\onescalar{1s}
\def\twoscalar{2s}
\def\withcontact{\times}

\newcommand{\PFinite}{P}
\newcommand{\Rat}{R}

\def\onehalf{{\textstyle\frac{\textrm{\small1}}{\lower2pt\hbox{\textrm{\small2}}}}}
\def\onequarter{{\textstyle\frac{\textrm{\small1}}{\lower2pt\hbox{\textrm{\small4}}}}}

\definecolor{allOrderBlue}{rgb}{0.4,0.5,1}
\definecolor{patternBlue}{rgb}{0,0,1}
\definecolor{photonRed}{rgb}{1,0.2,0.2}

\def\Ord{\mathcal{O}}

\def\backdexpatch latexntA{\hspace*{-6mm}}
\def\nsand#1.#2.#3{%
  \left\langle\smash{#1}{\vphantom1}\right|{#2}%
  \left|\smash{#3}{\vphantom1}\right]}
\def\nsandaa#1.#2.#3{%
  \left\langle\smash{#1}{\vphantom1}\right|{#2}%
  \left|\smash{#3}{\vphantom1}\right\rangle}
\def\nsandbb#1.#2.#3{%
  \left[\smash{#1}{\vphantom1}\right|{#2}%
  \left|\smash{#3}{\vphantom1}\right]}
\def\nsandba#1.#2.#3{%
  \left[\smash{#1}{\vphantom1}\right|{#2}%
  \left|\smash{#3}{\vphantom1}\right\rangle}

\def\spa#1.#2{\left\langle#1\,#2\right\rangle}
\def\spb#1.#2{\left[#1\,#2\right]}

\def\spash#1.#2{\spa{\smash{#1}}.{\smash{#2}}}
\def\spbsh#1.#2{\spb{\smash{#1}}.{\smash{#2}}}

\def\fig#1{Fig.~{\ref{#1}}}

\def\figs#1#2{Figs.~{\ref{#1}} and {\ref{#2}}}

\def\eqn#1{eq.~(\ref{#1})}
\def\eqns#1#2{eqs.~(\ref{#1}) and (\ref{#2})}

\def\<{\langle}
\def\>{\rangl\noaffiliatione}

\graphicspath{{./figures/}}

\title{{\small SAGEX-22-31-E, MITP-22-072}\\Yang–Mills All-Plus: Two Loops for the Price of One}
\ShortTitle{Yang–Mills All-Plus: Two Loops for the Price of One}

\author[a]{David A. Kosower}
\author*[b]{Sebastian P\"ogel}

\affiliation[a]{Institut de Physique Th\'eorique, CEA, CNRS, Universit\'e Paris--Saclay,\\
  F--91191 Gif-sur-Yvette cedex, France}

\affiliation[b]{PRISMA Cluster of Excellence, Institut f{\"u}r Physik, Johannes Gutenberg-Universit{\"a}t Mainz,\\ 
D--55099 Mainz, Germany}

\emailAdd{david.kosower@ipht.fr}
\emailAdd{poegel@uni-mainz.de}

\abstract{We present work  
on two-loop amplitudes in pure Yang--Mills theory 
with all gluons of
identical helicity. We show how to obtain
their rational terms --- the hardest parts to compute ---  
via well-understood one-loop unitarity techniques.}

\FullConference{%
  Loops and Legs in Quantum Field Theory - LL2022,\\
  25-30 April, 2022\\
  Ettal, Germany
}

\begin{document}
\maketitle

\section{Introduction}

Increasing integrated luminosity at the Large Hadron Collider 
(LHC) in the coming decade will drive experimenters' search for 
physics beyond the Standard Model (SM).
The LHC will be sensitive to ever-fainter discrepancies from SM 
predictions, thanks to 
increased statistics and to a better understanding of systematic 
uncertainties.  Greater experimental sensitivity does not suffice.
We also need higher-precision calculations in
perturbative QCD to reduce theoretical uncertainties.

The current frontier for perturbative QCD calculations is at
next-to-next-to-leading order (NNLO), where one expects
a reduction in these latter uncertainties to below a few percent.

We explore a technique for computing certain contributions to a
simple class of two-loop Yang--Mills amplitudes, the 
so-called ``all-plus''
amplitudes, with all external gluons of identical helicity.
This technique is based on a conjecture by 
Badger, Mogull and Peraro (BMP) in ref.~\cite{Badger:2016ozq}.
It relies solely on well understood one-loop generalized unitarity 
technology, and turns out to apply to all partial amplitudes in the 
color decomposition, in particular
also to the nonplanar ones.  We have given a much more detailed
discussion of our calculations in ref.~\cite{Kosower:2022bfv}.

\section{All-Plus Amplitudes}

All-plus amplitudes are simpler than general two-loop amplitudes.
At tree level, they vanish.
At one-loop, the dimensionally regulated amplitude ($D=4-2\eps$)
can be decomposed in a form exposing its universal singular 
structure~\cite{Giele:1991vf,BDDK,Kunszt:1994np,Catani:1996vz},
\begin{equation}\label{eq:one-loop_catani}
  \AOne = \ATree I^{(1)} + F^{(1)} + \mathcal{O}(\epsilon)\,.
\end{equation}
Here, $I^{(1)}$ is a universal function of the Lorentz invariants with double 
and single poles in $\eps$.
The vanishing of the tree makes one-loop amplitudes free of 
ultraviolet and infrared
divergences. Indeed the latter are purely rational in the external
spinors~\cite{Ellis:1985er,Bern:1993mq,Bern:1993qk,Mahlon:1992fs,Bern:2005hs,Mahlon:1993si}.
For the leading-color partial amplitude,
an all-$n$ conjecture
ref.~\cite{Bern:1993qk} came from demanding correct collinear factorization,
\begin{equation}\label{eq:A1_n:1_all-plus}
  \AOne (1^+\mathellipsis n^+)=
  -\frac{1}{3}\frac{\sum_{1\le i < j < k < l \le n}\spab{i|jkl|i}}
  {\spaa{12}\spaa{23}\mathellipsis\spaa{(n-1)n}\spaa{n1}}+\mathcal{O}(\epsilon)\,,
\end{equation}
a form which was later proven in ref.~\cite{Mahlon:1993si},
and rederived in ref.~\cite{Bern:2005hs}.
The subleading-color amplitudes at one-loop can always be obtained from the
leading-color ones through color relations~\cite{Bern:1990ux}.
Compact forms for them are also known~\cite{Bern:1993qk,Dunbar:2019fcq}.
Their finiteness and absence of branch-cuts
makes these expressions more like tree-level amplitudes than
one-loop ones.

The four-point all-plus amplitude at two loops
was computed long ago in ref.~\cite{Bern:2000dn};
the five-point one was derived much later~\cite{Badger:2013gxa,Badger:2015lda,Gehrmann:2015bfy,%
Abreu:2017hqn,Badger:2019djh}.
The full six-gluon amplitude was derived in 
refs.~\cite{Dalgleish:2020mof}, and there exist partial 
seven-gluon~\cite{Dunbar:2017nfy} and $n$-gluon 
expressions~\cite{Dunbar:2020wdh}.
More recently, first results for the four-gluon amplitude at three 
loops have been derived in refs.~\cite{Jin:2019nya,Caola:2021izf}.

As in the one-loop case, two-loop amplitudes can be decomposed with
respect to their singularity structure~\cite{Catani:1998bh}, which
leads us to a
relation similar to that of eq.~\eqref{eq:one-loop_catani},
\begin{equation}\label{eq:IR_Catani}
  \ATwo= \ATree I^{(2)} + \AOne I^{(1)} + F^{(2)} + \mathcal{O}(\epsilon).
\end{equation}
Here, $I^{(2)}$ is (like $I^{(1)}$) a universal function of the Lorentz invariants
with divergences up to $\eps^{-4}$;  $I^{(1)}$ is the same function as 
in eq.~\eqref{eq:one-loop_catani}.
The remainder $F^{(2)}$ is finite in dimensional
regularization, and can be split into polylogarithmic and
rational parts $P^{(2)}$ and $R^{(2)}$,
\begin{equation}
  F^{(2)}=P^{(2)}+R^{(2)}\,.
\end{equation}
Two-loop all-plus amplitudes have 
singularities in dimensional regularization of
the same degree as general one-loop amplitudes.
The polylogarithmic part $P^{(2)}$ has ordinary
branch cuts, and may therefore be computed using four-dimensional generalized unitarity.
In contrast, the rational part $R^{(2)}$ does not contain such
discontinuities and requires separate treatment.

The structure of the all-plus amplitude at two loops has made it possible to compute the polylogarithmic terms for an
arbitrary number of external gluons for the leading-color~\cite{Dunbar:2016cxp} and
a special subleading-color partial amplitude~\cite{Dunbar:2020wdh}.
In addition, the authors of refs.~\cite{Dunbar:2016gjb,Dunbar:2016aux,Dunbar:2016cxp,%
Dunbar:2017nfy,Dalgleish:2020mof,Dunbar:2019fcq} 
used recursive techniques to compute the rational terms in the 
five- and six-point amplitudes at 
leading and subleading color, as well as the leading-color 
seven-point amplitude~\cite{Dunbar:2017nfy}.
Dunbar, Perkins, and Strong~(DPS) presented an all-$n$
conjecture~\cite{Dunbar:2020wdh} for the special subleading-color 
amplitude.
BMP~\cite{Badger:2016ozq} 
have also computed the
leading-color five- and six-point rational parts through a reconstruction of the
integrand.
In addition, they presented a conjecture for the all-$n$
integrand at leading color on which we shall rely in our calculations.

\section{Separability}
In dimensional regularization, scattering amplitudes generally depend on two types of 
dimensional parameter: the dimension of loop-momentum 
integrations, $D$; and $D_s$, which controls
the number of states, with $D_s\geq D$.
Integrands of loop amplitudes depend \emph{polynomially} 
on $D_s$, while
integrals depend in a general analytic fashion on $D$.
The conjecture of BMP~\cite{Badger:2016ozq} is given in terms of the all-plus amplitude's 
dependence on the dimension $D_s$.

To extract this dependence, 
we follow a modified approach originally introduced in
ref.~\cite{Bern:2000dn}, and later exploited at one loop in
ref.~\cite{Giele:2008ve} and at two loops by 
BMP~\cite{Badger:2016ozq} to help isolate rational
contributions.
An all-loop discussion can be found in ref.~\cite{AccettulliHuber:2019abj}.

Any two-loop amplitude $A_{D_s}^{(2)}$
in Yang--Mills theory can be written as a quadratic
polynomial in $D_s$. 
By computing the amplitude for three different (ideally integer) values of $D_s$, 
we can fix the coefficients of this polynomial, allowing us to interpolate to 
non-integer $D_s$.
We refer to this method as \emph{dimensional reconstruction}.

Choosing $D_s=6,7,8$ as sampling dimensions, we can determine $A_{D_s}^{(2)}$~\cite{AccettulliHuber:2019abj},
\begin{equation}
  \label{eq:A2L_dim_reconstruction}
  A_{D_s}^{(2)}=A_{0}^{(2)}+(D_s-6)A_{\onescalar}^{(2)}+(D_s-6)^2
  A_{\twoscalar}^{(2)}+(D_s-6)(D_s-5)A_{\withcontact}^{(2)}\,.
\end{equation}
Here, $A_{0}^{(2)}$, $A_{\onescalar}^{(2)}$, $A_{\twoscalar}^{(2)}$, and $A_{\withcontact}^{(2)}$
are six-dimensional amplitudes with either two gluons ($A_{0}^{(2)}$), one gluon and one scalar ($A_{\onescalar}^{(2)}$), or two scalars running in the loops ($A_{\twoscalar}^{(2)}$, $A_{\withcontact}^{(2)}$).
The difference between $A_{\twoscalar}^{(2)}$ and $A_{\withcontact}^{(2)}$ lies in the way the two scalar loops are connected. In the former, they interact through the exchange of a gluon, while in the latter they are connected by a four-scalar contact term~\cite{AccettulliHuber:2019abj}.

In six dimensions, there are four different gluon polarization 
states. 
The scalars arise via Kaluza--Klein reduction from the additional 
polarization states of seven- and eight-dimensional gluons, and are 
massless in six dimensions.

In ref.~\cite{Badger:2016ozq}, BMP show how to decompose the leading-color two-loop
all-plus amplitudes into polylogarithmic terms $P_{n:1}^{(2)}$
and rational parts $\Rat_{n:1}^{(2)}$ associated to
different powers of the state dimension $D_s$.  
More precisely, through $\Ord(\eps^0)$,
BMP conjectured that these terms are always associated to different powers of $D_s-2$,
\begin{equation}
  F^{(2)}_{n:1}(1^+\mathellipsis n^+)=
  \onehalf (D_s-2)\PFinite_{n:1}^{(2)}(1^+\mathellipsis n^+)
  +\onequarter (D_s-2)^2\Rat_{n:1}^{(2)}(1^+\mathellipsis n^+)+\mathcal{O}(\eps)\,.
  \label{eq:two-loop_finite_statement}
\end{equation}
BMP verified this decomposition for the five- and six-gluon leading
color partial amplitudes.

Assuming this conjecture, we can use dimensional reconstruction to 
express the leading-color rational part 
in terms of six-dimensional amplitudes' rational parts.
Comparing \eqns{eq:A2L_dim_reconstruction}{eq:two-loop_finite_statement},
\begin{equation}
  \begin{aligned}
    \RT_{n:1}(1^+\ldots n^+) &= 4\bigl[\Rat_{\twoscalar}^{(2)}+\Rat_{\withcontact}^{(2)}\bigr]\equiv 4\RT_{ss}\,,
  \end{aligned}
  \label{eq:back_sep_PR_dimensional_reconstruction}
\end{equation}
where $\RT_{\twoscalar}$, $\RT_{\withcontact}$ are the rational parts of $\ATwo_{\twoscalar}$ and
$\ATwo_{\withcontact}$, and $\RT_{ss}$ is a shorthand for their sum.

Phrasing the BMP conjecture in the dimensional reconstruction picture  means we only need
the rational part $\RT_{ss}$.
The scalar Feynman rules forbid diagrams 
with propagators carrying both loop momenta.
All contributing two-loop Feynman integrals then
factorize into a product of one-loop integrals.
We can then determine the two-loop rational parts $\RT_{ss}$---and therefore $\RT(1^+\ldots n^+)$---using only
one-loop generalized unitarity techniques, in what we call the \emph{separable approach}.

We wish to compute the rational part $\RT_{ss}$
using one-loop $D$-dimensional generalized unitarity.
As all integrals have to factorize, we can limit ourselves to a basis of integrals which factorize as well.
Box, triangle and bubble integrals form a basis
of one-loop Feynman integrals. 
A basis of factorizing two-loop integrals is therefore
given by all integrals of the following six topologies,
\begin{equation}
    \foreach \x in {1,...,6}{%
    \includegraphics[valign=c,height=1.5em]{separability/int_classes/graph_\x}\hspace{0.5em}
  }\,.
\end{equation}

\begin{figure}
    \centering
    \includegraphics[width=5cm]{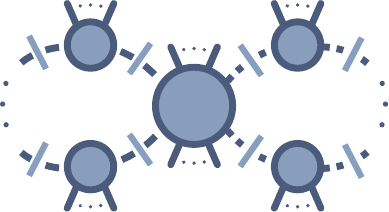}
    \caption{The generic of one-loop squared cuts contributing to $\RT_{ss}$.}
    \label{fig:generic_squared_cut}
\end{figure}

We use generalized unitarity cuts in
each loop to determine the coefficients of these types of 
integrals, following refs.~\cite{Forde:2007mi,Badger:2008cm}.
A generic ``one-loop squared'' cut contributing to the leading-color rational part is shown in \fig{fig:generic_squared_cut}.
The dashed lines represent the internal six-dimensional scalar 
loop propagators that are cut, while the circles are six-dimensional, color-ordered, on-shell tree amplitudes.
To compute the coefficient of each integral, we treat the loops 
sequentially.  The first loop is
computed from tree amplitudes using standard
one-loop unitarity techniques; the second loop is computed
from tree amplitudes and the coefficient computed at
the first step, with the latter playing the role of
a tree amplitude.  In the second step,
we again use one-loop unitarity.
As the loops are equivalent we are free to choose, which loop is 
computed first, and which one second.
The coefficient of the required two-loop integral is then 
the result of this two-stage computation.

The BMP conjecture leads to the separable approach for 
leading-color amplitudes~\cite{Badger:2016ozq}.
We extend the conjecture to the rational contributions
of subleading-color partial amplitudes as well, specifically the 
nonplanar ones.
That is, the nonplanar rational part $\RT_{ss}$ is related to the 
nonplanar two-loop rational parts of all-plus amplitudes
as in \eqn{eq:back_sep_PR_dimensional_reconstruction}.
Furthermore, thanks to the factorization of the loop integrals, we 
only require one-loop generalized-unitarity technology.

\section{Cuts for Color Structures}

We compute $\RT_{ss}$ for nonplanar amplitudes via color-dressed 
unitarity, similar to the procedure for
polylogarithmic contributions in refs.~\cite{Dunbar:2019fcq,Dunbar:2020wdh,Dalgleish:2020mof}.
We can write a complete 
two-loop $\text{SU}(N_c)$ Yang--Mills amplitude as
follows~\cite{Dunbar:2019fcq,Dalgleish:2020mof},
\begin{equation} 
\scalebox{0.9}{$
  \begin{aligned}
    \mathcal{A}^{(2)}_n &=
    N_c^2 \sum_{\sigma\in \modout{S_n}{Z_n}}
    \TTr\bigl(\sigma(1\mathellipsis n)\bigr)
    \,A^{(2)}_{n:1}\bigl(\sigma(1\mathellipsis n)\bigr)
    \\
    & +N_c \sum_{r=3}^{\lfloor \sufrac{n}{2}\rfloor +1} \sum_{\sigma\in \modout{S_{n}}{P_{n:r}}}
    \TTr\bigl(\sigma(1\mathellipsis(r-1))\bigr)\TTr\bigl(\sigma(r\mathellipsis n)\bigr)
A^{(2)}_{n:r}\bigl(\sigma(1\mathellipsis(r-1);r\mathellipsis n)\bigr)
    \\
    &+\sum_{r=2}^{\lfloor \sufrac{n}{2}\rfloor}\sum_{k=r}^{\lfloor \sufrac{(n-r)}{2}\rfloor}
    \sum_{\sigma\in\modout{S_{n}}{P_{n:r,k}}}\TTr\bigl(\sigma(1\mathellipsis r)\bigr)
    \TTr\bigl(\sigma((r+1)\mathellipsis(r+k))\bigr)
    \TTr\bigl(\sigma((r+k+1)\mathellipsis n)\bigr)
    \\
    &\hspace{12em}\times A^{(2)}_{n:r,k}\bigl(\sigma(1\mathellipsis r;(r+1)\mathellipsis(r+k);
    (r+k+1)\mathellipsis n)\bigr)\\
    &+\sum_{\sigma\in \modout{S_n}{Z_n}}\TTr\bigl(\sigma(1\mathellipsis n)\bigr)
    A^{(2)}_{n:\oneB}\bigl(\sigma(1\mathellipsis n)\bigr)\,,
  \end{aligned}
  $}
  \label{eq:two-loop_color_decomposition}
\end{equation}
where the $P_{n:r}$ and $P_{n:r,k}$ account for exchanges of the traces, as well as cyclic permutations of their arguments. 
The $\TTr(\ldots)$ represent traces over the color generators.
Interpreting factors of $N_c$ as empty traces, we can identify two distinct classes of color structures: those
with three traces, associated to $\ATwo_{n:1}$, $\ATwo_{n:r}$,$\ATwo_{n:r,k}$, and those with just a single color trace, associated to $\ATwo_{n:\oneB}$.

We can give a stringy heuristic argument for these two classes.
Two-loop gauge theory amplitudes can be obtained from the 
infinite-tension limit of genus-two open-string amplitudes.
External particles are realized through operator insertions on the boundary of the world-sheet.
The gauge group is introduced by dressing these insertions with Chan--Paton factors (in this case color generators), which are contracted along the boundary.

\begin{figure}[tbh]
\centering
  \captionsetup[subfigure]{justification=centering}
  \begin{subfigure}[]{0.49\linewidth}
    \centering
    \includegraphics[valign=c,height=1.1cm]
    {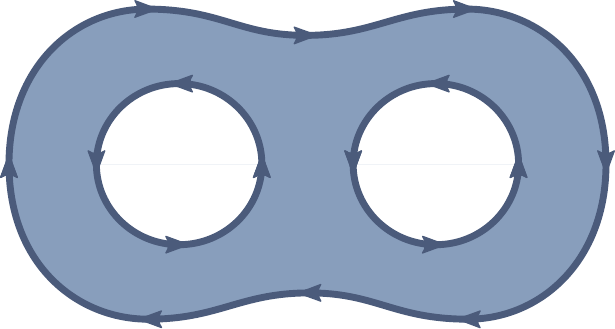}
    \caption{}
    \label{fig:ws_triple_trace}
  \end{subfigure}
  \begin{subfigure}[]{0.49\linewidth}
    \centering
    \includegraphics[trim={0 0 0 1.2cm},clip,valign=c,height=1.1cm]
    {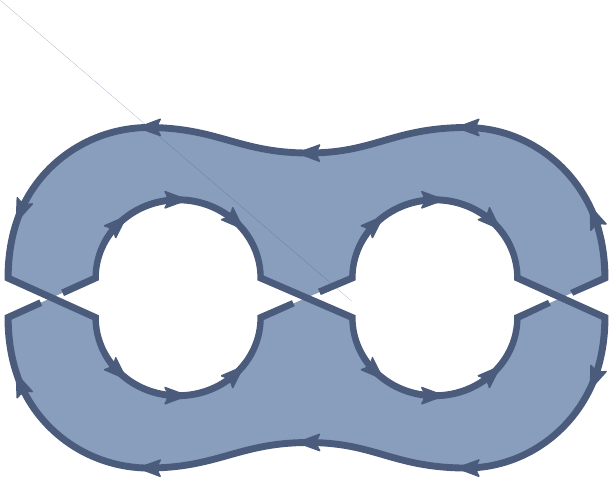}
    \caption{}
    \label{fig:ws_single_trace}
  \end{subfigure}
  \caption{The two types of genus two surfaces contributing to open-string amplitudes. 
  The one on the left has three boundaries, and generates three color traces.
  The one on the right has only a single boundary, therefore generating only one such trace.}
  \label{fig:world_sheets}
\end{figure}

Two types of genus-two surfaces contribute:
the kind shown \fig{fig:ws_triple_trace} has three boundaries, while the type shown in \fig{fig:ws_single_trace} has only a single boundary.
The former generates three color traces, and can be interpreted as the origin of partial amplitudes 
$\ATwo_{n:1}$, $\ATwo_{n:r}$, $\ATwo_{n:r,k}$.
The latter can only generate a single trace, and is therefore associated to the $\ATwo_{n:\oneB}$ partial amplitudes.
This string theory picture leads to one-loop squared cuts for the 
different color structures.

\begin{figure}[tbh]
\centering
  \captionsetup[subfigure]{justification=centering}
  \begin{subfigure}[]{0.40\linewidth}
    \centering
    \includegraphics[height=2.9cm]
    {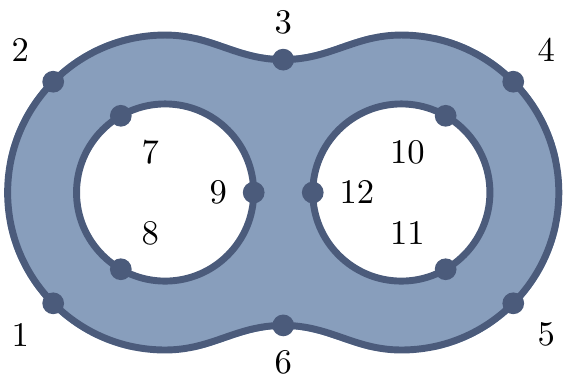}
    \caption{}
    \label{fig:cut_punctured_disk_topo}
  \end{subfigure}
  \begin{subfigure}[]{0.40\linewidth}
    \centering
    \includegraphics[height=3.4cm]
    {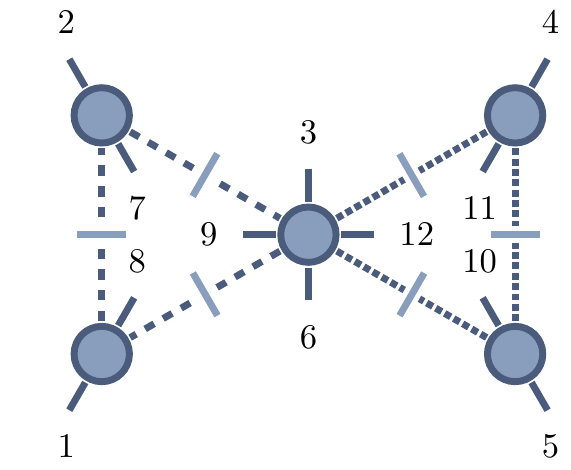}
    \caption{}
    \label{fig:cut_punctured_disk_cut}
  \end{subfigure}
  \captionsetup[subfigure]{justification=centering}
  \centering
  \begin{subfigure}{0.40\linewidth}
    \centering
    \includegraphics[height=8.8em]
    {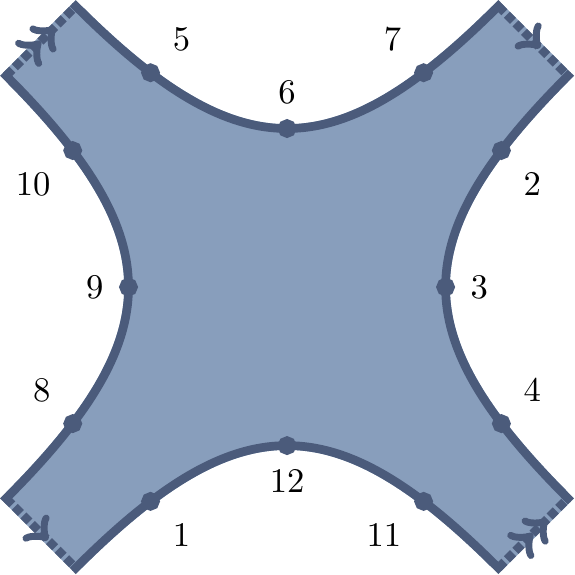}
    \caption{}
    \label{fig:1B_example_color_a}
  \end{subfigure}
  \begin{subfigure}{0.40\linewidth}
    \centering
    \includegraphics[height=8.8em]
    {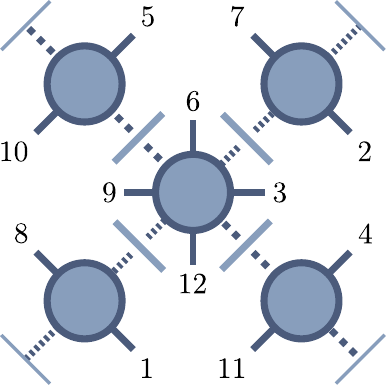}
    \caption{}
    \label{fig:1B_example_color_b}
  \end{subfigure}
  \caption{
    A graphical representation of the heuristic relation between
    the string world-sheet and unitarity cuts of subleading
    amplitudes.
    The configuration shown in (a) and (b) corresponds to the 
    color structure
    $\TTr(1, 2, 3 ,4, 5, 6) \TTr(7, 8, 9) \TTr(10,11,12)$.
    Figures (c) and (d) correspond to the single-trace
    structure $\TTr(1,2,\ldots,11,12)$. In the latter two, 
    the dotted lines need to be sewn together
    according to the arrows shown.}
  \label{fig:color_routing_example}
\end{figure}

To determine $\RT_{ss}$ for three-trace amplitudes, 
we use cuts of form shown
in \fig{fig:generic_squared_cut}.
We identify the outer and two inner edges of the cut with the three 
color traces, attaching
the external gluons to them accordingly.
This principle alongside the string theory motivation is 
illustrated in \figs{fig:cut_punctured_disk_topo}{fig:cut_punctured_disk_cut}.
We can check independently that dressing the color-ordered 
tree-amplitudes with their color traces and some algebra leads to
the correct trace structure.
The full rational part $\RT_{ss}$ is then the sum over unique cuts, taking into account all associations of traces to the three edges.

To obtain $\RT_{ss}$ for subleading single-trace amplitudes, we 
have to find cuts generated by
the single-edge surface of \fig{fig:ws_single_trace}.
We can smoothly deform this surface to an equivalent form, shown 
in \fig{fig:1B_example_color_a}. 
From this it is easy to identify the associated unitarity cuts, 
see \fig{fig:1B_example_color_b}. 
The key difference from the cuts for three traces is the attachment 
of the scalar lines: while before the two lines are separated at 
the connecting tree amplitude, they now cross.
The subleading single-trace rational part $\RT_{ss}$ is then 
given by the sum over all unique cuts of this form.
\section{Verification}
We verify the separability conjecture using an automated
generation of nonplanar cuts in \mm.
Their evaluation is carried out using the one-loop $D$-dimensional unitarity techniques presented in refs.~\cite{Forde:2007mi,Badger:2008cm}.
This procedure is also automated in \mm\ using a custom series expansion code specialized for
the type of rational functions that appear.
The code is capable of handling numerical (rational), as well 
as analytically parametrized, kinematics.
All numeric comparisons were done exactly on rational 
kinematic points%
\footnote{The current version of the \mm\ packages used for the generation and evaluation of cuts can be found  \href{https://github.com/spoegel/SpinorHelicityPackages}{here}.}.

Using our code we find exact numerical agreement of the separable 
construction with 
all known analytic results in the literature.
We compared the rational parts of all four, five, and six-gluon 
partial amplitudes of 
refs.~\cite{Dunbar:2016gjb,Dunbar:2016aux,Dunbar:2016cxp,%
Dunbar:2017nfy,Dalgleish:2020mof,Dunbar:2019fcq}, as well as the 
leading color seven-gluon amplitude of ref.~\cite{Dunbar:2017nfy}.
Using kinematics based on parametrized momentum twistors, we also 
rederived closed analytic forms for
all five-gluon rational terms.
We further verified numerically up to nine gluons the 
agreement with the conjecture for 
the rational parts of the subleading single-trace 
amplitudes~\cite{Dunbar:2020wdh}.
 
\section{Conclusions}

In this proceeding, we have explored rational terms in 
two-loop amplitudes.  These terms exhibit the least
symmetry or structure
of all contributions to Yang--Mills amplitudes.
We studied the simplest such terms,
in the all-plus gluon amplitudes.
We relied on the BMP separability conjecture and
used a straightforward extension of generalized unitarity
techniques for computing one-loop rational terms in order
to compute them.  We computed the rational
terms in the four- and five-point two-loop amplitudes
analytically, and those in the six- and seven-point
amplitudes numerically.  Our results agree with
those obtained by Dunbar, Dalgleish, Jehu, Perkins and Strong through a recursive
approach.  They also agree numerically
with the DPS all-$n$ conjecture~\cite{Dunbar:2020wdh}
for the subleading-color single-trace 
amplitude at eight and nine points.  In addition to
evidence for the correctness of the results
in refs.~\cite{Dunbar:2017nfy,Dalgleish:2020mof,Dunbar:2019fcq}, 
our calculations
also provide evidence for the correctness of the separability
conjecture~\cite{Badger:2016ozq} both for 
leading- and subleading-color
amplitudes.  
The ideas developed here may also help simplify
the calculation of other rational terms at two loops,
in particular in the other simple helicity configuration,
with a lone negative-helicity gluon.

\section{Acknowledgements}
We would like to thank the organisers of Loops and Legs for a stimulating conference.
The research described here
has received funding from the European Union's 
Horizon~2020 research and innovation program
under the Marie Sk\l{}odowska-Curie grant agreement
No.~764850 ``SAGEX''.
DAK's work was supported in part by the French 
\textit{Agence 
Nationale pour la
 Recherche\/}, under grant ANR--17--CE31--0001--01,
 and in part by the European Research Council, under
 grant ERC--AdG--885414. 
SP's work has been supported in part by the Mainz Institute for Theoretical Physics (MITP) of the Cluster of Excellence PRISMA+.

\end{document}